# Stray light and polarimetry considerations for the COSMO K-Coronagraph

Alfred G. de Wijn[*a], Joan T. Burkepile[a], Steven Tomczyk[a], Peter G. Nelson[b], Pei Huang[c], and Dennis Gallagher[a]

[a]High Altitude Observatory, National Center for Atmospheric Research, P.O. Box 3000, Boulder, CO 80303, USA; [b]Sierra Scientific Solutions, 1540 Patton Drive, Boulder, CO, USA; [c]Consultant to NCAR, 1484 N. Larkspur Ct., Lafayette, CO, USA

## ABSTRACT

The COSMO K-Coronagraph is scheduled to replace the aging Mk4 K-Coronameter at the Mauna Loa Solar Observatory of the National Center for Atmospheric Research in 2013. We present briefly the science objectives and derived requirements, and the optical design. We single out two topics for more in-depth discussion: stray light, and performance of the camera and polarimeter.

## 1. INTRODUCTION

The solar corona is a million-degree plasma that is the source of the solar wind and the site of explosive activity such as Coronal Mass Ejections (CMEs) that drive space weather throughout the heliosphere. The corona is organized by the Sun's magnetic field into brighter magnetically 'closed' regions, where the plasma is contained by the field, and magnetically 'open' regions of very low density where the plasma and field are carried outward to form the solar wind. Coronal brightness and the distribution of closed and open regions vary over the 11-year solar cycle. Dynamo processes in the solar interior determine the quantity and distribution of magnetic flux into the corona and drive the sunspot activity cycle.

Much of what is known about CME properties and the density structure of the corona comes from white-light observations.[1,2,3,4,5,6,7] These observations are needed to understand CME formation, as well as the evolution of the global structure of magnetic field and density distribution of the corona. They also provide insight into the mechanisms responsible for coronal heating and solar wind acceleration. The K-Coronameter at the Mauna Loa Solar Observatory (MLSO) has been operating since 1980 and provides unique white-light observations of the low corona. The K-Coronameter utilizes a 1-D CCD detector to acquire scans of the low corona every ½ degree in azimuth and builds up an image of the corona over 360 degrees every 3 minutes. The new K-Coronagraph is scheduled to replace this aging instrument at MLSO in 2013. It will significantly improve our understanding of the formation and dynamics of CMEs and the global density structure of the corona through improved sensitivity, cadence, and field of view (FOV).

Even at the best sites and on the best days, the foreground sky brightness drowns out the corona. However, because the white-light K-corona is caused by electron scattering of photospheric light, it is linearly polarized tangential to the solar limb. Ground-based coronagraphs for white-light measurements exploit this by making a measurement of polarization brightness (pB), which can be related to coronal electron density.

The K-Coronagraph is one of the three instruments in the proposed COSMO suite. The two other instruments of COSMO are the Chromosphere and Prominence Magnetometer (ChroMag) and the Large Coronagraph for coronal magnetic field measurements. ChroMag is currently in a prototyping phase, while the Large Coronagraph is in the preliminary design phase.

## 2. K-CORONAGRAPH SCIENCE OBJECTIVES AND REQUIREMENTS

CMEs are explosive events driven by magnetic stresses in the solar atmosphere and are the primary driver of space weather at earth. CMEs form and accelerate low in the corona with speeds and accelerations that can vary over 3 orders of magnitude. The greatest acceleration occurs below 3 $R_{sun}$ for most events. Observations of their onset and measurements of the rate-of-change of CME acceleration are needed to discriminate between the many models posited to explain their formation. This requires rapid image sequences of the very low corona with a FOV from 1.05 to 3.00 $R_{sun}$ to track CMEs from their formation in the first coronal scale height (1.01 to 1.09 $R_{sun}$) to the height where they have acquired most of their acceleration (3 $R_{sun}$). CME structure sets the required spatial sampling at 6 arcsec or better.

---

[*] dwijn@ucar.edu, phone +1 303 497 2171

Table 1. Instrument requirements for the COSMO K-Coronagraph.

| Quantity | Units | Requirement | Goal | Mk4 |
|---|---|---|---|---|
| FOV | $R_{sun}$ | 3 | 4 | 2.9 |
| Lower Limit of the FOV | arcsec | 50 | 25 | 120 |
| Spatial Sampling | arcsec | 6 | 3 | 5×9 to 5×23 |
| Noise Level | pB/√Hz | $3.9\times10^{-9}$ | $1.3\times10^{-9}$ | $5.4\times10^{-8}$ |
| Map Time | s | 15 | 8 | 180 |

Very high time cadence (13 to 40 s) images recorded by TRACE and YOHKOH SXT have allowed measurements of acceleration changes over very limited heights for a very small number of events due to their limited FOV.[8] A 15-s time cadence will provide the coverage that is needed to record accurate trajectories of the fastest CMEs.

Earth-directed "halo" CMEs are considerably fainter due to projection effects than those happening at the limb and ejected perpendicular to the line of sight. The noise level must be sufficiently low to detect coronal structures with intensities of $10^{-9}$ $B_{sun}$.

The instrument requirements from these objectives are summarized in Table 1.

## 3. OPTICAL DESIGN OVERVIEW

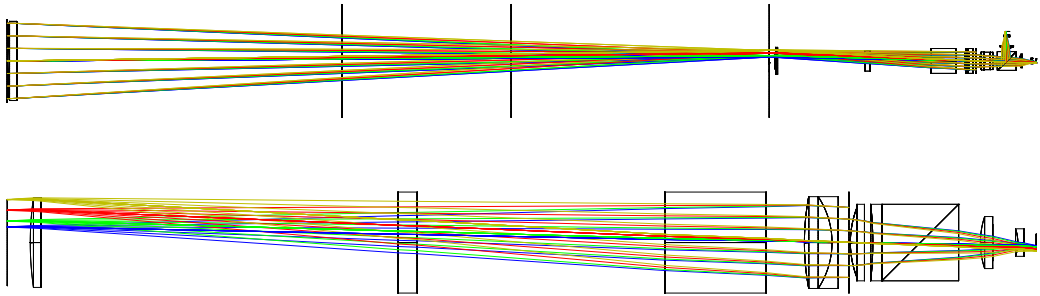

Figure 1. Overview of the COSMO K-Coronagraph optical design (top) and zoom-in on the back-end optics (bottom). The optical design shows the aperture stop and primary lens (O1) at the left and several baffles in the telescope tube. The zoom-in shows the prime focus (where the occulter is located) at the left, followed by the field lens, the bandpass filter, the modulator stack (here shown as a monolithic block), a doublet lens to correct O1 chromatic aberration, the Lyot stop, the first two elements of the camera lens, the polarimetric analyzer polarizing beam splitter, the last two elements of the camera lens, and finally the focal plane at the far right. The total optical track is 2734 mm.

Figure 1 shows an overview of the optical design as well as a more detailed view of the compact back-end. The instrument design meets or exceeds the requirements given in Table 1. The COSMO K-Coronagraph is a traditional Lyot coronagraph with a 20-cm fused silica primary (O1), a field lens just behind the occulter at prime focus, and a Lyot stop at the image of the aperture. The bandpass filter, the polarimetric modulator (in this model shown as a monolithic block), and a doublet lens that corrects the chromatic aberration of the primary are placed between the field lens and the Lyot stop. The first two elements of the camera lens system are placed immediately after the Lyot stop, followed by the polarizing beamsplitter analyzer for the polarimeter, and finally the last two elements of the camera lens. The O1 has focus capabilities in order to make a sharp image of the sun at the occulter. The camera focal planes move parallel to the optical axis to bring the occulter in focus.

The K-Coronagraph is a so-called dual-beam polarimeter, i.e., the modulated light is analyzed in two perpendicular directions and recorded simultaneously using two synchronized cameras. This technique allows for the elimination of nearly all crosstalk from intensity to polarization resulting from seeing, pointing jitter, etc. The camera lens system consists of four optical elements. The polarizing beamsplitter is placed in between the second and third optic, so that the first two elements are shared between the cameras.

## 4. STRAY LIGHT ANALYSIS

There are four main contributors to scattered light in any Lyot-type coronagraph: diffraction, ghosting, O1 surface roughness, and O1 surface contamination.

### 4.1 Diffraction

Light from the solar disk is diffracted by the telescope aperture, and some of the diffracted light spills over the occulter. The image at prime focus is truncated by both the occulter and the field stop. As a result, at the image of the aperture created by the field lens the diffracted light is concentrated in a ring at the edge of the aperture and at a spot in the center. The Lyot Stop blocks most of this light (as well as any light scattered off the edge of the aperture). The reduction in occulter edge diffraction by the Lyot Stop is given by:

$$\text{LS Attenuation} = \frac{P \times \Delta\beta^2}{2\pi(1-P)\frac{d}{0.001\lambda}} \times \left\{ \frac{1}{\left[\left(\frac{m \times s}{7200"} + \Delta\beta\right)\left(\frac{m \times s}{7200"}\right)\right]^2} + \frac{1}{\left[\left(\frac{(2+m) \times s}{7200"} + \Delta\beta\right)\left(\frac{(2+m) \times s}{7200"}\right)\right]^2} \right\}^2 ,$$

where $P$ is the fractional projected area of Lyot Stop, and in the case of the K-Coronagraph, $m = 0.02$ (occulter over sizing in units of the solar disk), $d = 200$ mm (O1 diameter), $l = 0.735$ mm (center wavelength), $s = 1920$ arcsec (avg. solar diameter), and $Db$ = angular distance from occulter edge (rad). In order to provide sufficient attenuation yet not block more light than necessary we chose $P = 0.85$.

### 4.2 Ghosting

The main risk of ghosting is associated with the double bounce inside the O1 lens since 1) it is uncoated; and 2) the solar disk light is very bright. A strong attenuation of this ghost by geometrical defocus is needed in order to not contribute significantly to the 2 ppm near-disk irradiance attenuation goal at the occulter. The solar disk ghost image at the occulter is about 137 cm in diameter compared to 1.78 cm for the main image. Due to the reflectance of uncoated fused silica (3.4% at 725 nm), we have an intensity attenuation of $1.2 \times 10^{-3}$, resulting in a total ghost level of $\sim 1 \times 10^{-10}$ of that of the main disk at the detector, well below the required stray light levels.

### 4.3 Surface Roughness Analysis

The micro-roughness requirement is derived from the requirement that the integrated scattered light observed 48 arc seconds above the solar limb (0.05 radii) should be less than 2 ppm of the solar disk center. 2 ppm is approximately one third of the best scattered light level expected with a clean lens at this height, but ~4 times the brightness of the sky at MLSO under ideal sky conditions.

We select the common "ABC model" (also known as the K-correlation model) as a basis for our calculations:

$$S_2(f) = \frac{A_2}{\left[1 + (B f)^2\right]^{(C+1)/2}}.$$

The subscript $_2$ refers to the fact that this is a model for a 2-D power spectrum, following the convention in Stover.[9] Note that the 1-D spectrum has the same form, except the power in the denominator is $C/2$; the slope of the 2-D spectrum is one power of frequency steeper than the 1-D spectrum of the same surface, at high spatial frequencies. In this expression, therefore, $C$ is the slope of the 1-D power spectral density (PSD). $A_2$ gives an overall scaling and $B$ describes at what frequency below which the power law becomes flat.

Among the best-characterized sets of fused silica optics of the appropriate dimension are those made for the Laser Interferometer Gravitational-wave Observatory. Their ⌀250×100mm mirrors were polished to ~5 Å RMS over the frequency band of 4.3–7500 cm$^{-1}$. They measured the PSD of their micro-roughness over a very wide frequency range from 0.1 to 10,000 cm$^{-1}$. Those data show the 1-D PSD to have a slope $C$ of ~1.3 with no evidence of a low-frequency flattening (described by the $B$ parameter). For our modeling we therefore set $C = 1.3$ and $B = 10^5$ μm. The scattered light level and the RMS micro-roughness (over a given band) then become only a function of the parameter $A_2$.

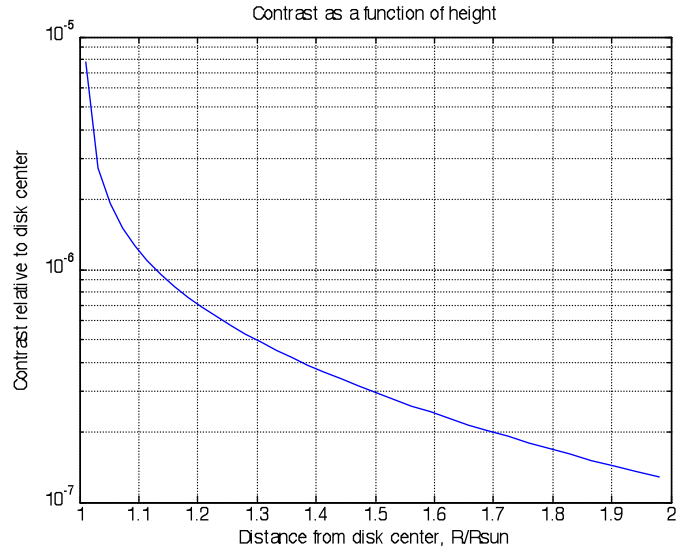

Figure 2. The predicted level of scattered light for a 5 Å RMS finish. The calculation includes solar limb darkening at 725 nm.

The value of $A_2$ is chosen such that the stray light requirement is met. The RMS surface roughness specification is then derived. To calculate the scattered light level we use the Modeled Integrated Scatter Tool (MIST) developed by the Optical Technology Division of the National Institute for Standards and Technology (NIST). We find that the spatial frequencies that contribute to the scattered light are from $3.23 \times 10^{-4}$ to $1.25 \times 10^{-2}$ μm$^{-1}$. In order to meet the 2 ppm requirement at 48 arcsec from the solar limb the RMS roughness over this bandwidth must be 5 Å or less.

Figure 3 shows the predicted level of scattered light for the specified surface finish quality. This calculation includes the effect of solar limb darkening at 725 nm. The 5 Å micro-roughness requirement is consistent with the requirements of other internally-occulted coronagraphs (such as the COR-1 instruments on the STEREO spacecrafts).

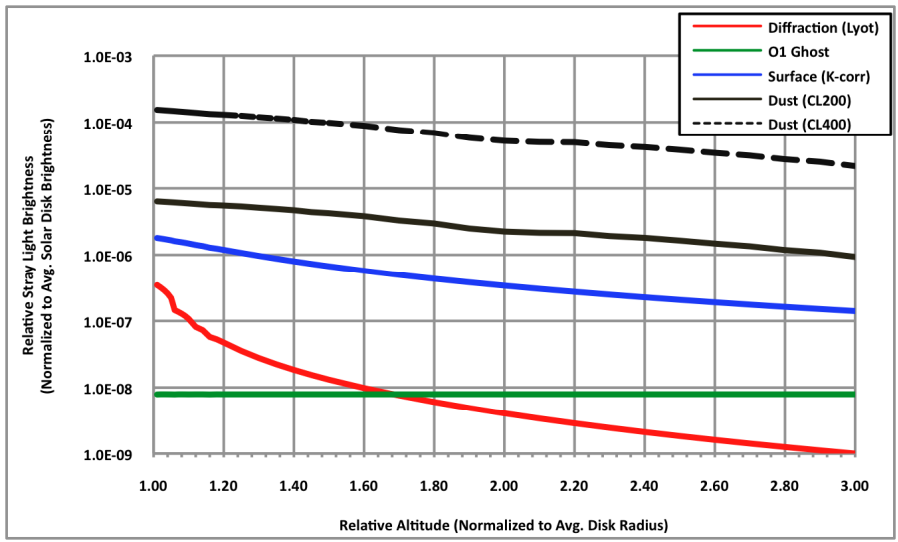

Figure 3. The predicted stray light contributions from diffraction, O1 ghost, surface roughness, and surface contamination.

### 4.4 Surface Contamination Analysis

Stray light caused by scattering from lens surface contamination was modeled at two levels, CL200 and CL400. This is a Mie scattering calculation result, which can be calculated using MIST or approximated.[10]

### 4.5 Summary

In order to make comparisons between the results of the stray light models above and real-world coronal measurements, we must integrate all of the Point Source Transfer (PST) functions to form an overall stray light irradiance figure for each of the effects, and then scale the result to the solar disk brightness. The result is shown in Figure 3.

Diffraction would dominate the stray light at the low range of the FOV in the absence of the Lyot stop, but for the baseline stop of 85% of the aperture area it becomes small relative to other contributors. The O1 ghost is also insignificant. While the O1 surface roughness is critical, it is clear that O1 surface contamination is the most important factor in ensuring low levels of scattered light. A HEPA system was incorporated in the design to keep the O1 bathed in filtered air in order to reduce the build-up of contamination. Despite these efforts it is expected that the O1 lens will have to be cleaned on a regular basis.

## 5. INSTRUMENT MODEL

The K-Coronagraph performs a noise-dominated measurement. As such, a high photon flux is desired so that averaging can be employed to improve the signal-to-noise. The figure of merit of the camera thus is photo-electrons per second, i.e., the product of the well depth and the frame rate. Most scientific cameras have shallow wells and/or slow frame rates. Some custom cameras exist, but they tend to be expensive and difficult to implement. We selected the PhotonFocus MV-D1024E-160-CL-12 CMOS machine-vision camera for its deep wells (180 ke$^-$) and high frame rate (135 Hz). This camera provides a 12-bit readout. In order to verify that this industrial camera can still provide science-quality output, a sophisticated instrument model was developed.

### 5.1 Instrument Model

A known Stokes vector is sent into the model. It is passed through the telescope, modulated, analyzed, and converted into an electron count in the camera. The electron count is then sampled using a model of the Analog-to-Digital Converter (ADC). 506 readouts (approximately corresponding to a 15-s observation) are accumulated in each of the four modulation states, then summed, demodulated and calibrated. The experiment is repeated 500 times to gather statistical information on the noise of the measurement. The resulting measured Stokes vectors are compared with the input Stokes vector.

The parameter space is scanned in intensity from 2 to $18 \times 10^{-6}$ $B_{sun}$ with a 7-ms exposure (though the results can be scaled to different intensity/exposure ranges), and in degree of Stokes Q polarization from 0 to 10%. These conditions cover excellent to poor sky conditions. We assume that no U or V polarization enters the telescope.

### 5.2 Digitization

ADCs in the camera convert electrons captured in the wells into a digital readout. The process of digitization introduces a systematic error. A measurement at a specific intensity nearly always returns the same discrete data number (DN) if the combination of shot and read noise is much smaller than the least significant bit (LSB). The actual average readout then differs from the expected average readout. In order to reach $10^{-4}$ accuracy with a 12-bit camera, we must be able to determine an average to better than 0.1 DN. A simple calculation shows this translates into a requirement that the noise must be at least 11 e$^-$ for the MV-D1024E, easily satisfied by the 220 e$^-$ read noise. Note that if the MV-D1024E was an 8-bit camera, the noise would have to have been 350 e$^-$; this increase in noise required to remove the systematic effects of discretization disqualifies most 8-bit cameras for our application.

### 5.3 Bit Errors

In a perfect ADC, the value of each bit is exactly half that of the next more significant bit. However, in practice the bits exhibit small variations in size that we will call "bit errors". Bit errors manifest themselves as systematic offsets in the demodulated measurements. The ADCs in an MV-D1024E camera were characterized in the lab. The histograms show some values are favored over others. This can introduce a systematic error that we model here. Using the measured histograms we simulate the ADC in our instrument model, and derive the systematic and random errors resulting from the measurement. After discussion with PhotonFocus the camera was modified with a drop-in replacement 14-bit ADC (reading out only the 12 most significant bits), which reduced the systematic error by about a factor of two. However, the error remains at the $10^{-3}$ level (Figure 4).

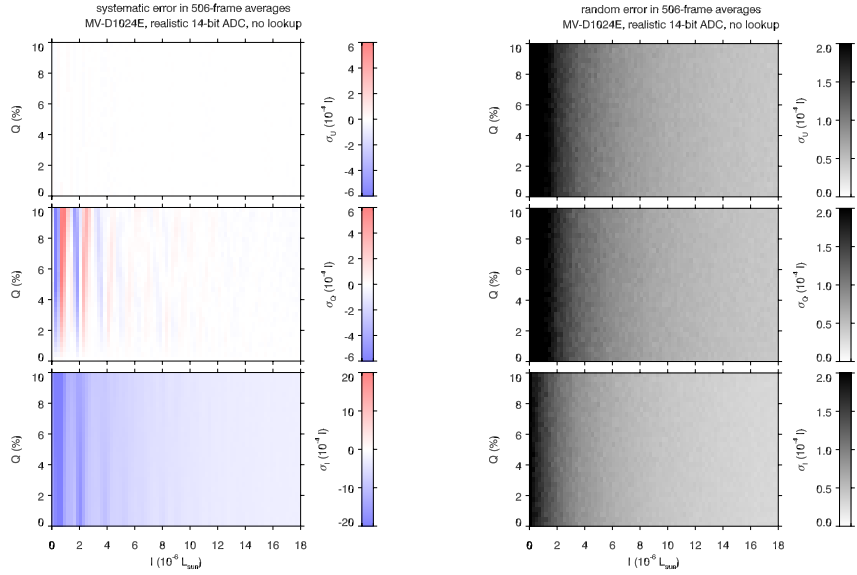

Figure 4. Systematic (left) and random (right) errors from a simulation that includes measured bit errors of an ADC in the MV-D1024E as a function of intensity and degree of polarization for a pure Stokes-Q signal. From top to bottom: error in Stokes U, Q, and I.

Further improvement is made by applying a lookup table to correct individual readouts. The histogram used to encode the ADC characteristics is derived from different pixels than the one used for the correction. The latter also has 0.5% normally distributed random noise added. The lookup table returns a 22-bit integer number, so that 1024 exposures can be added without risk of overflow in a 32-bit integer. Figure 5 shows that the errors are at the $10^{-4}$ level. Notice that systematic errors in $I$ are similar in size to the random errors. Since the errors are determined on the basis of 500 model evaluations, the statistical noise on the determination of the systematic error is $1/\sqrt{500} \approx 0.044$ of the random error. We conclude that the noise in the lookup table dominates the systematic errors.

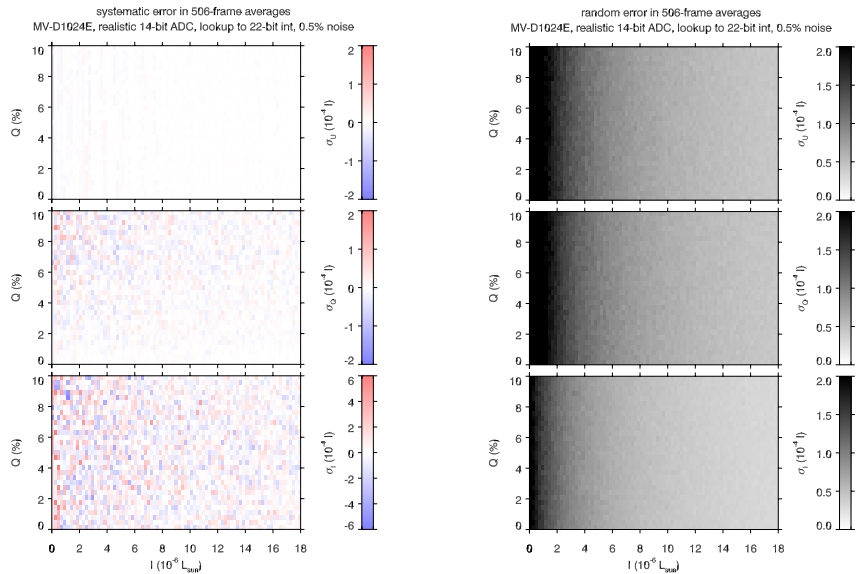

Figure 5. As Figure 4 but using a lookup table to correct for bit errors.

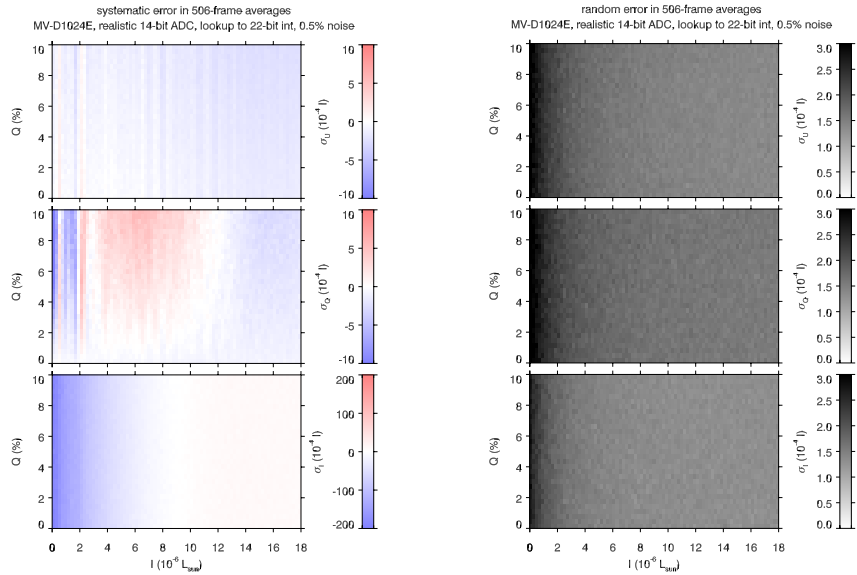

Figure 6. As Figure 5 but including sensor nonlinearity, dark current, a realistic modulation matrix, a telescope matrix derived from the ZEMAX model, and synthesized calibration.

## 5.4   Calibration

The above tests all employed a perfect modulation and calibration. The functioning of the polarimeter and calibration system must also be verified. The model was extended to include dark current and sensor nonlinearity. The dark current is given in the MV-D1024E data sheet; for the nonlinearity we use the 3rd order polynomial fit given in the EMVA 1288 Standard test of the PhotonFocus MV-D1024E-160-CL-12 camera by AEON Verlag & Studio. It is corrected by applying a 4th order polynomial inverse in the lookup table. Furthermore, the modulation matrix is based on a realistic design (and thus non-ideal), and the telescope matrix is taken from the ZEMAX model. The analyzer is assumed to be a 99.9% polarizer in one beam, 99% in the other, but otherwise perfect.

First, 50000 sets of calibration observations are synthesized. In this process the calibration polarizer is modeled as a 99.9% polarizer and a diffuser intensity of $10^{-5}$ $B_{sun}$ is used. These sets are then individually and independently processed to generate 50000 sets of calibration data consisting of the dark current, gain, and modulation matrix in each beam.

The calibration data are then used in a model run to calibrate the observations. Each model experiment uses a randomly chosen set of calibration data. The random noise is slightly increased compared to the perfect calibration (as can be expected). The systematic errors in Q are at the $10^{-3}$ level or below (Figure 6).

The systematic errors in I are large, at the $10^{-2}$ level. Since I is not a difference measurement, it is much more sensitive to errors in the determination of the dark current, gain, nonlinearity, etc. An absolute error of 1% is not unreasonable in real-world measurements. For the K-Coronagraph, I is not a measurement of interest, so this issue does not concern us.

## ACKNOWLEDGMENTS

The National Center for Atmospheric Research is sponsored by the National Science Foundation.